# Interplay between Young Stars and Molecular Clouds in the Ophiuchus Star-forming Complex

Aashish Gupta[1,2,3] and Wen-Ping Chen[1,4]
[1] Graduate Institute of Astronomy, National Central University, 300 Zhongda Road, Zhongli, Taoyuan 32001, Taiwan; Aashish.Gupta@eso.org
[2] Academia Sinica Institute of Astronomy and Astrophysics, No. 1, Sec. 4, Roosevelt Road, Taipei 10617, Taiwan
[3] European Southern Observatory, Karl-Schwarzschild-Str. 2, D-85748 Garching bei München, Germany
[4] Department of Physics, National Central University, 300 Zhongda Road, Zhongli, Taoyuan 32001, Taiwan

## Abstract

We present spatial and kinematic correlation between the young stellar population and the cloud clumps in the Ophiuchus star-forming region. The stellar sample consists of known young objects at various evolutionary stages, taken from the literature, some of which are diagnosed with Gaia EDR3 parallax and proper-motion measurements. The molecular gas is traced by the 850 $\mu$m Submillimetre Common-User Bolometer Array-2 image, reaching $\sim$2.3 mJy beam$^{-1}$, the deepest so far for the region, stacked from the James Clerk Maxwell Telescope/Transient program aiming to detect submillimeter outburst events. Our analysis indicates that the more evolved sources, namely the class II and III young stars, are located further away from clouds than class I and flat-spectrum sources that have ample circumstellar matter and are closely associated with natal clouds. Particularly the class II and III population is found to exhibit a structured spatial distribution indicative of passage of shock fronts from the nearby Sco–Cen OB association thereby compressing clouds to trigger star formation, with the latest starbirth episode occurring now in the densest cloud filaments. The young stars at all evolutionary stages share similar kinematics. This suggests that the stellar patterns trace the relics of parental cloud filaments that now have been dispersed.

*Unified Astronomy Thesaurus concepts:* Star formation (1569); Molecular clouds (1072); Young stellar objects (1834); Young star clusters (1833); Star forming regions (1565)

## 1. Introduction

It is likely that all stars are formed in groups out of giant molecular clouds. A star group may soon be dispersed upon emerging from the natal cloud (Lada & Lada 2003). Subsequent gas dispersal, two-body relaxation between member stars, and stellar ejection exacerbate the disassembly process, leaving behind a star cluster we are witnessing now. With time, external perturbations, such as tidal forces plus differential rotation, continue to dissolve the star cluster to supply the disk field stars.

While the formation of individual stars in observation or in theory is reasonably understood, particularly for low-mass stars, certain details of stellar birth remains unclear. For instance, are stars of different masses formed coevally? Do luminous members tend to be centrally located in a star cluster as their birthplace, or are they dynamically settled with time? How do these short-lived massive stars influence the subsequent star formation on nearby clouds? While dense molecular cores, within which individual stars form, appear to have already mass segregated (Kirk et al. 2016), tidal disruption or dynamical ejection could lead to certain spatial distribution of protostars and pre-main-sequence (PMS) stars versus cloud filaments (Stutz & Gould 2016). To address questions concerning, e.g., the intricate interplay of massive stars, molecular clouds, and young stars, a sample of the youngest stellar population recently exposed out of, but still closely associated with, molecular clouds is needed.

The Ophiuchus molecular cloud complex offers such a target. With the proximity of ~138 pc (Ortiz-León et al. 2018), it is one of the nearest known star-forming regions (Wilking et al. 2008) to render optimal sensitivity and angular resolution to diagnose the cloud morphology and, with a wealth of optical and infrared observations, the stellar kinematics. Particularly, the densest cloud core in the region, known as L1688, is considered the nearest site of cluster formation (Wilking & Lada 1983). The region is known to be influenced, or even with the starbirth triggered at least partly, by the nearby Sco–Cen OB (Sco OB2) association (see Wilking et al. 2008), with the cloud "streamers" as the leftover material of star formation (Vrba 1977). With an age spread of 0.3 Myr in the core (Chen et al. 1995; Wilking et al. 2008) to 2–4 Myr for the rest of L1688 (Esplin & Luhman 2020), the young stellar and substellar populations in L1688 represents a comparative sample at a variety of evolutionary stages (Dunham et al. 2015) from embedded protostars, PMS objects with dusty disks/envelopes, to relatively evolved PMS objects void of inner disks, and from aging massive stars to brown dwarfs in infancy, so as to shed light on the interaction and feedback between the cloud complex and the newborn stars.

This study uses the deepest 850 $\mu$m James Clerk Maxwell Telescope (JCMT)/Submillimetre Common-User Bolometer Array (SCUBA-2) image of L1688 to trace the cold dust, as a proxy of molecular cloud structure, so as to correlate spatially and kinematically with the sample of young stars in the region compiled from the literature (Wilking et al. 2008; Dunham et al. 2015; Allers & Liu 2020; Esplin & Luhman 2020), plus a few probable member candidates identified using the Gaia EDR3 parallax and proper-motion data (Gaia Collaboration et al. 2021). The PMS evolutionary status of these young stars, either already reported in the literature or quantified by us with archival photometric measurements across a range of wavelengths, then allows us to infer the star formation history in the region.

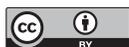

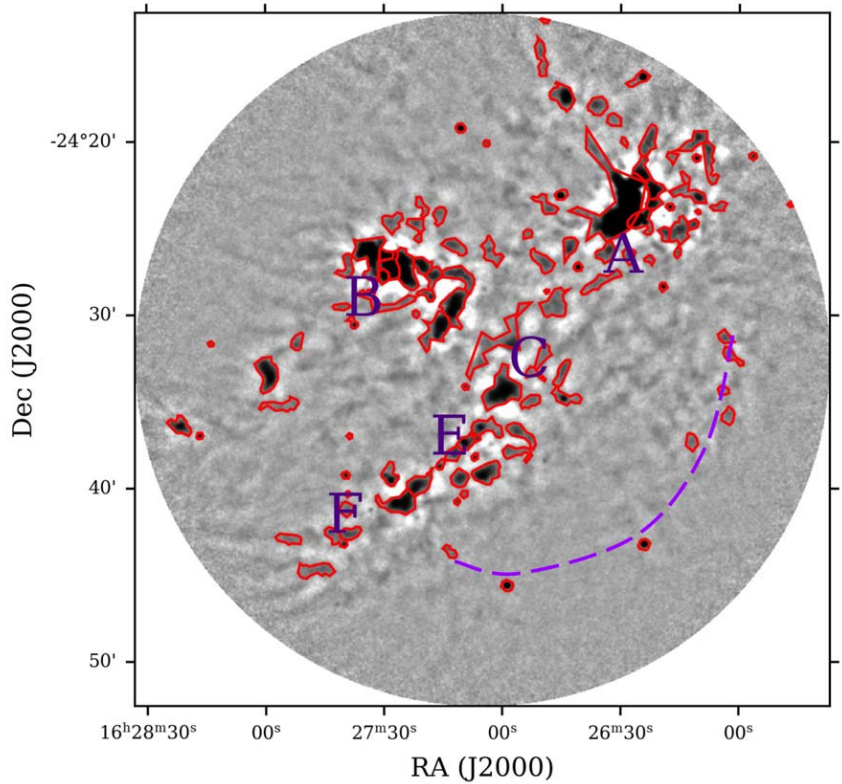

**Figure 1.** The JCMT/SCUBA2 850 $\mu$m emission map of L1688. The red polygons show the clump boundaries as identified by the FellWalker algorithm. Prominent clouds are labeled. Purple dashed line denotes the string of small clumps, as discussed in Section 3.

## 2. Data and Analysis

Data used in our study include submillimeter emission tracing cold dust in molecular clouds, and a sample of young stellar objects in L1688 collected from the literature. We are particularly interested in the spatial distribution and kinematics of young stars at different evolutionary stages relative to the cloud substructures.

### 2.1. JCMT 850 $\mu$m Clumps

To trace the molecular substructure, we used the 850 $\mu$m emission acquired with the SCUBA-2 (Holland et al. 2013) on the JCMT. SCUBA-2 is a 10,000 pixel bolometer taking continuum data simultaneously at 450 $\mu$m and 850 $\mu$m, with effective beam sizes of 9$''$8 and 14$''$6, respectively. The data adopted for this study were taken as a part of the JCMT Transient Survey (Herczeg et al. 2017; Johnstone et al. 2018), designed to detect continuum variability of young embedded sources utilizing multiepoch observations of star-forming regions, including L1688 in the Ophiuchus cloud. Raw data were reduced using MAKEMAP (Chapin et al. 2013), an iterative map making software. Maps of individual epochs were then stacked together, thereby creating a deeper coadded image (Mairs et al. 2017) for each region. For our study, we employed a coadd of 29 epochs of L1688 observations, reaching a sensitivity of $\sim$2.3 mJy beam$^{-1}$, about a factor of 5 lower than a typical single epoch observation, and roughly three times more sensitive than the JCMT Gould Belt Survey data (Pattle et al. 2015).

We identified enhanced submillimeter substructure, or "clumps" using the STARLINK implementation of the FellWalker algorithm (Berry 2015), which traces boundaries of clumps by following paths from low-valued to high-valued pixels in an iterative manner. Polygonal clump boundaries are identified and their peak fluxes and total fluxes are estimated. For our analysis, the key parameter for clump detection, the "noise," was set to be 10 mJy beam$^{-1}$. A total of 108 clumps were identified with the minimal peak flux being $\sim$30 mJy beam$^{-1}$.

Our study focuses on the spatial and kinematic distributions of YSOs relative to cloud clumps; no attempt was made to derive the temperature and mass of individual clumps. Figure 1 exhibits the latest coadded SCUBA-2 850 $\mu$m image of L1688, marked with clump boundaries. Our JCMT data are relevant to those reported by Johnstone et al. (2000) and by Pattle et al. (2015). Johnstone et al. (2000) used SCUBA, the predecessor of SCUBA-2, to cover a 700 arcmin$^2$ region, and with the *clfind* clump-finding algorithm (Williams et al. 1994), identified 55 clumps. These authors derived the clump mass by two models, one adopting a single temperature of 20 K for every clump, and the other by assuming the clump to be an isothermal, self-gravitating, pressure-confined Bonnor–Ebert sphere. The resultant clump mass functions are similar. Pattle et al. (2015) using SCUBA-2 to analyze L1688 along with parts of L1689, L1709, and L1712, identified 70 clumps in L1688 with the *CuTEx* algorithm (Molinari et al. 2011). Combining the 160 and 250 $\mu$m data from the Herschel mission (Ladjelate et al. 2020), plus the 450 $\mu$m data from SCUBA-2, Pattle et al. (2015) constructed the spectral energy distribution (SED) of each clump to estimate its temperature, and hence to derive the mass. The mass function of the clumps from a few solar masses down to ∼0.1 $M_\odot$ is found to have a slope consistent with that of the stellar initial mass function.

### 2.2. Young Stellar Objects

Our young stellar sample consists of a compilation of literature lists. The primary reference is the review article by Wilking et al. (2008), in which 238 objects are within the sky area of our study on the basis of stellar youth characteristics, such as infrared excess, lithium absorption, or X-ray emission. In terms of decreasing level of infrared excess as a proxy of PMS evolutionary sequence, from the youngest, embedded protostellar stage, through a transition phase, to the T Tauri stage, and finally to the inner-disk free stage, there are 23 class I, 28 flat-spectrum, 92 class II, and 36 class III sources. There are 59 sources without classification.

In addition, Dunham et al. (2015) identified young stars in the star-forming regions in the Gould Belt, including Ophiuchus, using data from the Spitzer Telescope (Werner et al. 2004), and found 14 additional sources, eight of which are class I or flat-spectrum sources. Toward lower masses, Allers & Liu (2020) reported the young substellar objects in Ophiuchus and Perseus, seven of which (four newly found) are within our field of view. Esplin & Luhman (2020) compiled a list of YSOs in the Ophiuchus cloud complex on the basis of spectroscopic evidence of youth; among these, 13 additional sources are included in our list.

We combined all these 269 sources as our young stellar sample. Of these, a total of 99 have Gaia EDR3 counterparts, each within a matching radius of 5″, with astrometric and photometric measurements down to the Gaia brightness limit of $g \approx 21$ mag. Figure 2 presents the Gaia EDR3 parallax and proper-motion measurements. In each case, the young sample stands out distinctly. However, there are 10 sources (GSS 39, Elia 2−31, V* V2675 Oph, EM* SR 24A, ISO-Oph 9, DROXO 106, EM* SR 23, VSSG 22, BBRCG 54, BKLT J162748−242204) found not to be in the parallax range; i.e., they are likely foreground or background objects. Incidentally, we recognized four sources coexisting and comoving with our YSO sample. Two of them (BKLT J162749−242522, BKLT J162818−242836) are known PMS stars, whereas the nature of other two (BKLT J162553−244129, BKLT J162702−241639) needs to be further vindicated; but, we included them in the member sample for now. At the end, we have a total of 263 YSO sources for analysis.

Cánovas et al. (2019) and Grasser et al. (2021) conducted a similar search for young stars in the whole Ophiuchus cloud complex using Gaia DR2 and EDR3 data, respectively. They also identified BKLT J162702−241639 and BKLT J162818−242836 as cluster members. All the new sources they identified as candidate young stars within our field of L1688 are already in our final list. All the YSOs have Gaia EDR3 parallax consistent with membership. Because the distance estimate as the direct reciprocal of the parallax introduces statistical bias (Bailer-Jones et al. 2021), this work focuses only on the projected sky location of stars with no consideration of the distance distribution.

#### 2.2.1. Kinematic Data

In order to investigate the kinematic properties of the cluster, we gathered the tangential velocities (proper motions) and radial velocities of the YSOs, when available. The Gaia EDR3 provides proper motions of 84 YSOs in our sample, incomplete for embedded sources. A literature search using Simbad led to 32 additional sources having proper-motion data in Ducourant et al. (2017) and one in Monet et al. (2003). The list in Ducourant et al. (2017), though greatly increasing the proper-motion data in our sample, is limited mainly to the Ophiuchus F core and partially to the E core. For radial velocities, the Gaia EDR3 data are available for only six objects, whereas Sullivan et al. (2019) provide data for 31 sources, and Torres et al. (2006) for additional one source.

#### 2.2.2. Evolutionary Classification

Dunham et al. (2015) reported the evolutionary status of the YSOs in the region. We conducted an independent analysis so as to compare with the previous results and also to expand the sample to include sources with no classification in the literature. We compiled for each source its SED as widely covered in wavelength as possible from archival photometric data. These include Pan-STARRS (optical Chambers et al. 2016), Gaia (optical), 2MASS (near-infrared; Skrutskie et al. 2006), WISE (mid-infrared; Wright et al. 2010), Spitzer IRAC and MIPS (mid- to far-infrared; Werner et al. 2004), Herschel (far-infrared; Pilbratt et al. 2010), and 450 and 850 $\mu$m fluxes (this work on the SCUBA-2 coadded maps, computed using aperture photometry with 10″ and 14″ apertures, respectively). For each SED, we computed the slope of the infrared spectral index, defined as $\alpha = d \log(\lambda S_\lambda)/d \log \lambda$, from wavelength 2 to 25 $\mu$m to classify an object's evolutionary state in terms of its infrared excess as the proxy of the amount of circumstellar dust, as per the following criteria (Dunham et al. 2014): (1) $\alpha \geqslant 0.3$ (class 0 to I); (2) $-0.3 \leqslant \alpha < 0.3$ (flat-spectrum sources); (3) $-1.6 \leqslant \alpha < -0.3$ (class II); (4) $\alpha < -1.6$ (class III).

In our analysis, 30 SEDs have inferior fits ($\sigma_\alpha > 0.5$) mainly arising from imperfect cross-matching among database catalogs or a lack of data in certain wavelength ranges, so were not included in further analysis. Even though our $\alpha$ index is not extinction corrected, of the 132 sources common with Dunham et al. (2015), 80% have consistent classifications, which is considered satisfactory given the systematic uncertainties due to inclination or aspherical geometries of the disks/envelopes, etc. For the few sources we did not have our own classifications, we adopted the classes from Dunham et al. (2015) or Wilking et al. (2008). At the end, our YSO sample consists of 30 Class 0/I sources, 56 flat-spectrum sources, 110 Class II sources, and 58 Class III sources, of which ∼21% of the sources had no classification prior to our analysis. Some ∼3% sources still remain unclassified.

## 3. Results and Discussion

In addition to distinct cloud clumps, the deep JCMT submillimeter data presented as Figure 1 reveal clearly the known filamentary morphology of the clouds, with cloud Oph-B manifest the "streamers" shaped by the passing shocks from the OB stars to the west and southwest, including but not limited to HD 147889 (B2 III), $\sigma$ Sco (O9.5 V + B7 V), and $\alpha$ Sco (M0.5 Iab + B3 V). The extension to the northeast immediately from Oph-A may be a part of the ring/shell created by $\rho$ Oph (B2 IV + B2 V). All these luminous stars are outside the field of our JCMT data and the readers may refer to Wilking et al. (2008) and

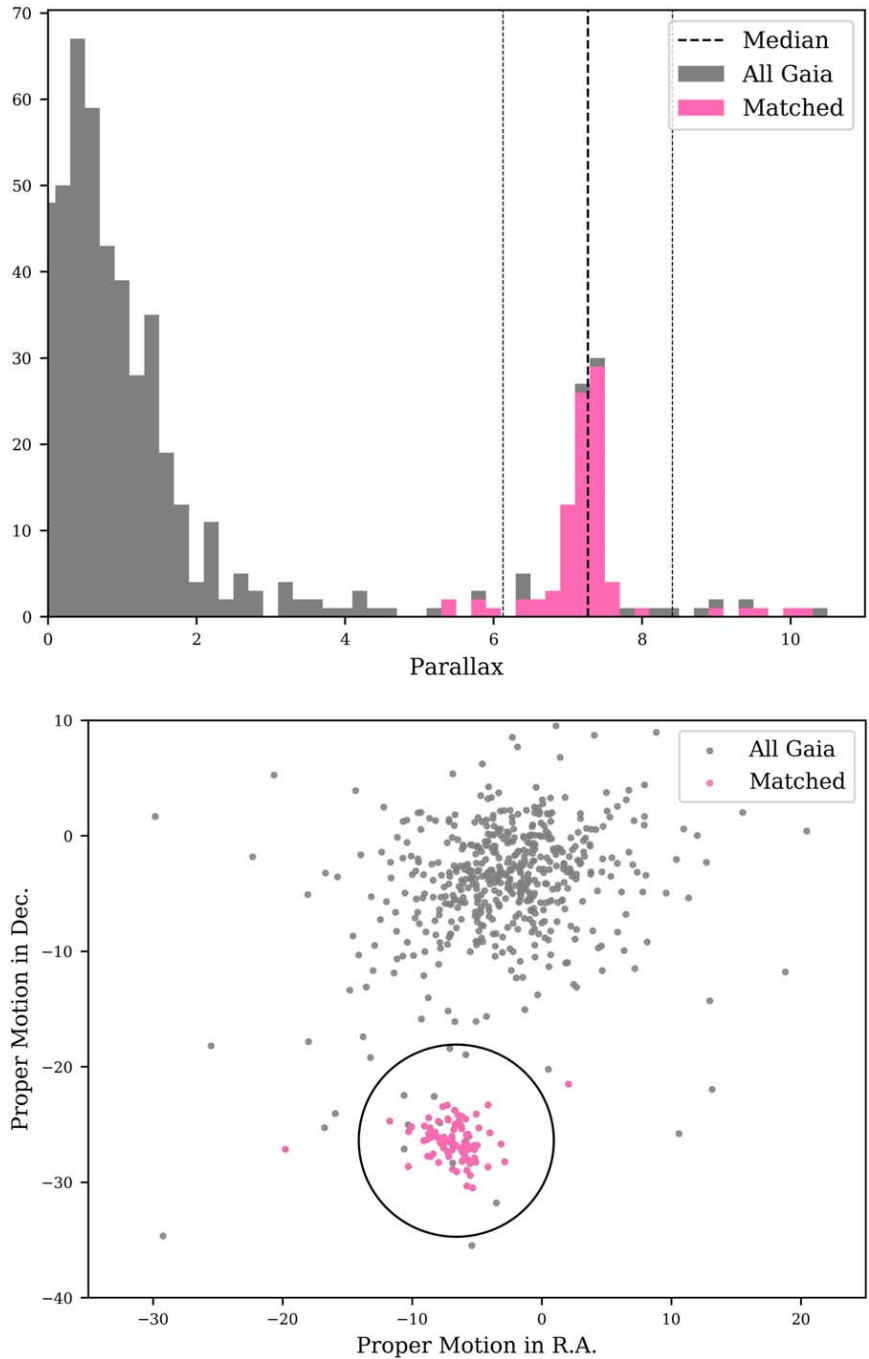

**Figure 2.** Gaia EDR3 data for (a) the parallax distribution, and (b) the vector plot of the proper motions. In each case, the gray symbol represents every Gaia star in the region, whereas the pink symbol marks known YSOs. In (a), the thick and thin dashed lines mark, respectively, the median and interval of five times median-absolute deviation. In (b), the ellipse is the region of five times the median-absolute deviation.

Nozawa et al. (1991), both with a $^{13}$CO map, or Ladjelate et al. (2020) with a Herschel map for a larger-scale illustration of the complex.

Motte et al. (1998) noticed the alignment of protostars along clouds Oph-C, Oph-E, and Oph-F. This direction of clump filaments (see also Johnstone et al. 2000), orthogonal to the steamers, arches to A, which probably acts as the "hub," i,.e., the conjunction of cloud filaments, along which mass flows to seed the formation of massive stars or star clusters (Kumar et al. 2021). The scenario is consistent with the magnetic field direction, inferred by optical and infrared polarization (Vrba et al. 1976; Vrba 1977) as a consequence of dichroic extinction due to grain alignment, that by and large parallels the dark cloud filaments.

The deep JCMT/SCUBA-2 image uncovers additional small clumps, for the first time to our knowledge, to the southwest of the main filaments A-C-E-F, forming a pattern that seems to loop between Oph-A and Oph-E/F, stretching to R.A. $\sim 16^h26^m$, decl. $\sim -24°32'$ where it is recognized previously as a part of cloud Oph-J (Simpson et al. 2008). This string of discrete cloud clumps is marked as the purple dashed line in Figures 1 and 3 and will be discussed later in the context of the YSO distribution.

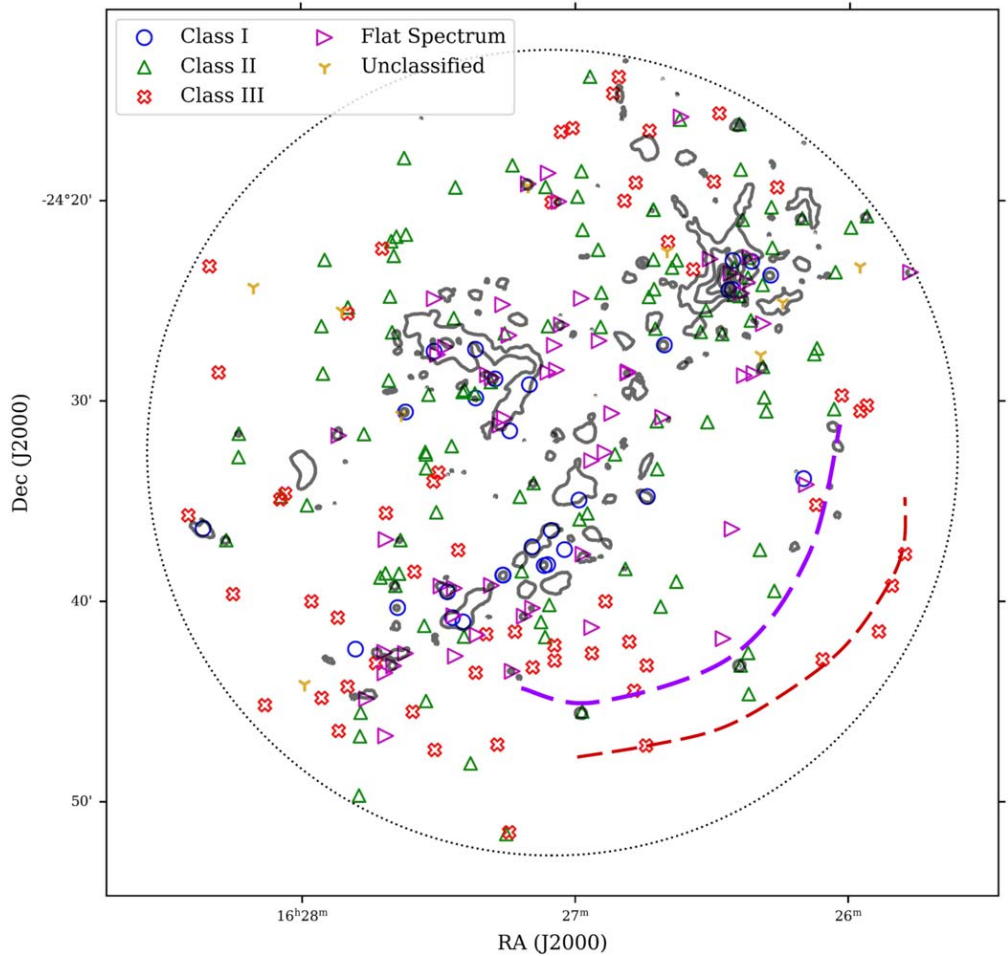

**Figure 3.** Spatial distribution of YSOs and of the 850 μm emission outlined by the contours. The YSOs (class I marked by blue circles, flat-spectrum by purple triangles, class II by green triangles, class III by red crosses, and unclassified sources by yellow symbols) are not randomly distributed. Instead, they appear to follow stream-like patterns. Inner purple dashed line denotes the string of small clumps, as discussed in Section 3 and illustrated in Figure 1. Outer red dashed line denotes string of evolved sources, as discussed in Section 3.1.

### 3.1. Stellar Spatial Distribution

Figure 3 displays the spatial distribution of the YSOs with respect to the 850 μm emission. It is seen that the class I and flat-spectrum sources are preferentially nearer the molecular clumped structure than more evolved sources are. This is consistent with the previous investigation in the same region by observations (Jørgensen et al. 2008), and by numerical simulation (Frimann et al. 2016).

The result is quantified in Figure 4 for which the projected linear separations of YSOs to the center of the nearest 850 μm clump, adopting a distance of 138 pc to the complex, are presented for different evolutionary classes. While more evolved sources are scattered equally likely up to 0.2 pc from a cloud, objects in infancy are an order closer in the vicinity of clouds, mostly within 0.02 pc. Also exhibited is the same separation distribution but for the sample of Dunham et al. (2015), which is smaller in size than ours but with reddening corrections. The distributions are similar for class O/I and flat-spectrum sources, but contrast mostly for the class III samples. Conceivably, a highly reddened class III object could be mistaken with an $\alpha$ index as one with significant IR excess. But the reverse is not true; that is, a genuine class I/O source would not have been recognized as an evolved class III source even without dereddening. The conclusion that less evolved YSOs, either with our sample or with that of Dunham et al. (2015), are physically more associated with dense gas clumps therefore remains valid.

Mairs et al. (2016) and Stutz & Gould (2016) found similar trends of younger stars being more closely associated with molecular clouds in the Orion A star-forming region, roughly 10 times in size scales of our field. Under the assumption that the YSOs were formed at the location of the presently observed cloud clumps, these authors attributed this trend to greater relative velocities of more evolved sources so as to being decoupled from parental molecular gas.

While clouds in L1688 have a hub-filamentary morphology, the young stars in the region also exhibit a structured spatial distribution. As discussed earlier, the class I and flat-spectrum sources are clustered around the cloud emission and thus follow the distribution of the clouds. The class II/III sources, though correlated with generally weaker gas emissions, appear to form arc-like patterns. Most prominent is the western partial loop between cloud Oph-A and clouds Oph-E/F connecting the 850 μm clumps (marked as a purple dashed line in Figure 3) that are associated with mostly flat-spectrum and class II sources. Further to the southwest as the outermost arc (marked as red dashed line in Figure 3) near the edge of our JCMT field are class III young stars with no corresponding submillimeter emission, thereby to our knowledge not recognized before.

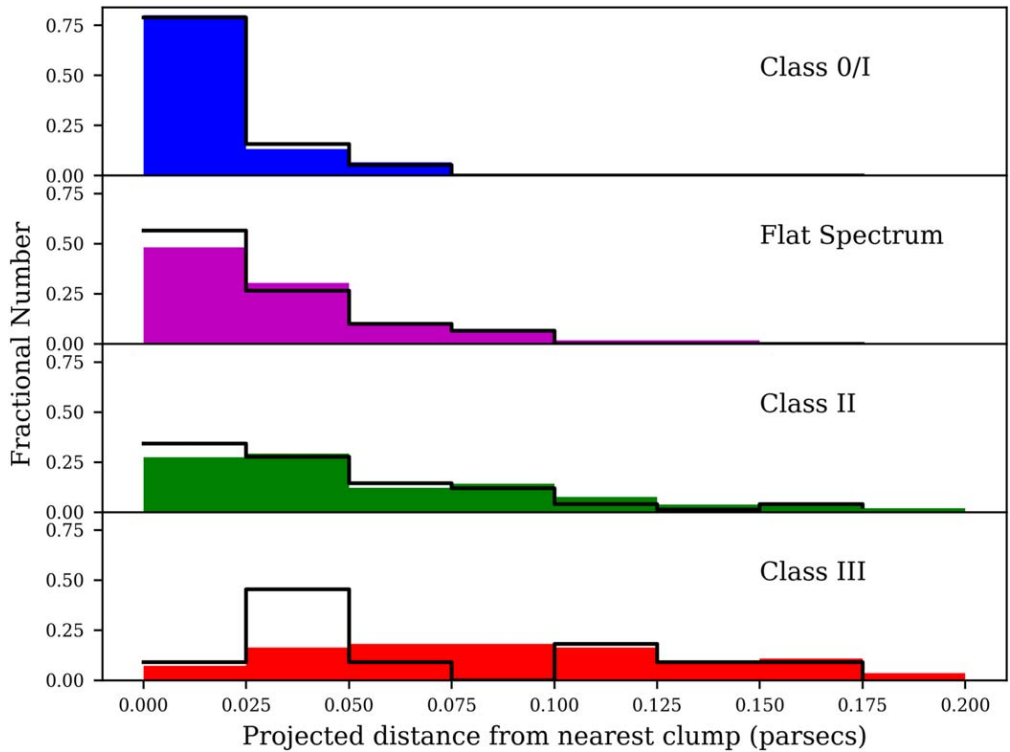

**Figure 4.** Projected distances from the nearest 850 μm emission clumps of YSOs at different evolutionary phases. The solid color bars show the distribution of the YSOs of different classes. The black lines show the distribution for the sample of Dunham et al. (2015).

To the other side of the main cloud filaments (A-C-E-F), between Oph-B and Oph-F additional stellar patterns appear to stand out, approximately in the NS of NE-SW direction. It is not clear whether these are parts of the cloud streamers or of the southwest YSO loops which are physically closer in 3d to the OB stars.

Such organized patterns could not have been the outcome of random ejection from common sites of formation. In order to test the randomness in the spatial distribution of the evolved sources, we used *pointpats* implementation of Ripley's K function (Ripley 1976), which is a proxy of percentage of points (in this case class II/III sources) that are within a certain distance of each other. It is commonly used to quantify spatial randomness for discrete location data, i.e., applicable to our case to address whether YSOs are distributed randomly or in a clustered pattern (see Retter et al. 2019). Figure 5 compares the K function of our YSO sample against that of an otherwise complete spatial randomness via Monte Carlo simulation. The observed K function lies outside the envelope at all angular scales considered here (less than 3′.5), leading to to a *p* value of 0.0001, thereby rejecting the null hypothesis of being a random distribution.

### 3.2. Stellar Kinematic Distribution

With a heliocentric distance of 138 pc, the Galactocentric coordinates of the region are $(X, Y, Z) = (-7990.23, -16.02, 60.26)$ pc, and from the median proper motion of $(\mu_\alpha \cos\delta, \mu_\delta) = (7.03, 26.01)$ mas yr$^{-1}$ and the median radial velocity of RV $= -7.33$ km s$^{-1}$, the space motion of the young stellar group is $(U, V, W) = (6.21, 230.43, -1.79)$ km s$^{-1}$.

We now present the motion of the YSOs relative to clouds based on available kinematic data, as discussed in Section 2.2.1. The peculiar motion of YSOs, i.e., the proper motion of individual sources subtracted by the median proper motion, is shown in Figure 6. Using these relative velocities, we inspected the kinematic distribution for different groups of YSOs.

Figure 7 presents the tangential velocity distributions of YSOs at different evolutionary stages. Lacking proper-motion data for highly embedded sources, we combined Class 0/I sources and flat-spectrum sources together. One sees that these least evolved population has a spread of velocities, partly perhaps because of the uncertainties in the data, whereas the more evolved sources, class II and class III alike, consistently move slowly. This supports the notion that the different YSO classes have similar velocities, and more evolved sources are further away from the clouds not because they are moving faster.

Motte et al. (1998) presented evidence, with the cloud morphology, kinematics, and magnetic field configuration, of multiepoch triggered star formation in ρ Oph by a combination of plausible mechanisms, such as the shocks from Sco–Cen OB association by supernova explosion, massive stellar winds, or ionization fronts.

The two southwestern arcs (or partial layers in 3d) of YSOs reported in this work provide morphological evidence of the direction of the star-forming sequence propagating from the southwest now to the dense segment extending from A to F. Kinematic analysis indicates that all the YSOs remain near their formation groups and hence retain the spatial structures of their natal clouds. The earlier episode of star formation to the southwest is imprinted in the YSOs as relic of the filamentary or sheet-like clouds. The impact of massive stars occurred at an earlier epoch on the chain of class III sources that are more evolved or depleted much of the circumstellar disks and interstellar cloud.

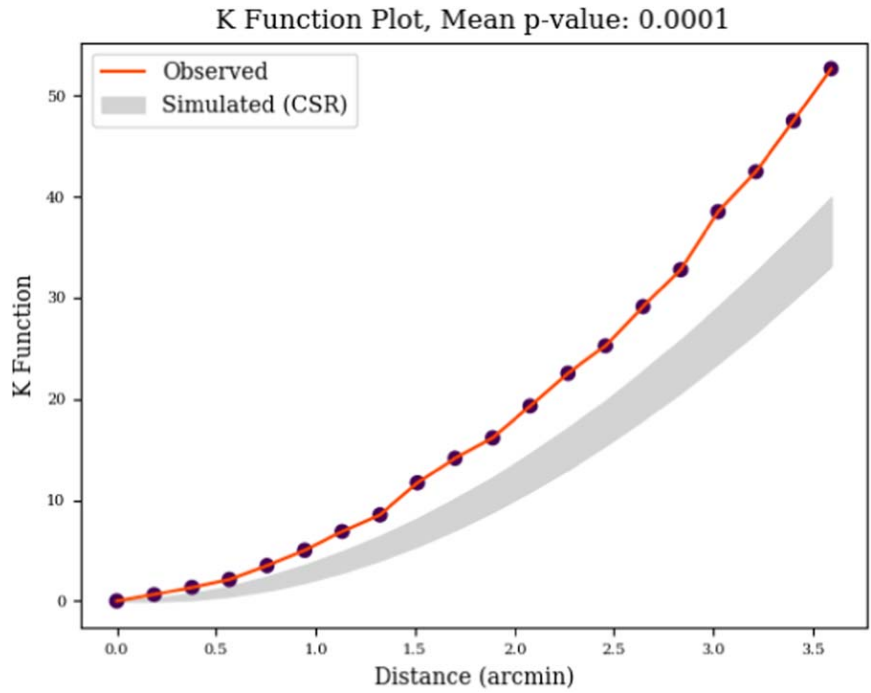

**Figure 5.** The Ripley's K function of the class II and III objects in our sample (in red) is clearly different from that of central 95 percentile interval (the shaded gray region) of the K function of a randomly distributed simulated sample.

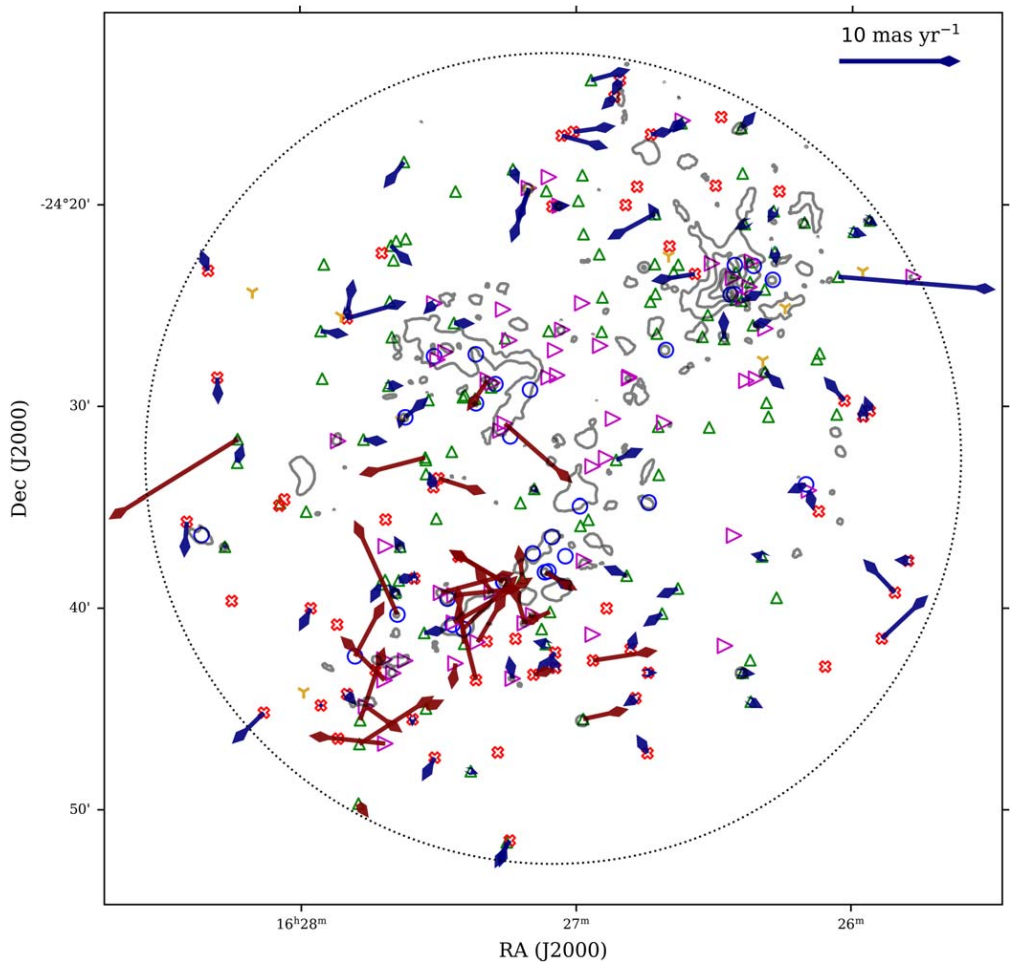

**Figure 6.** The peculiar motions of YSOs, i.e., the proper motion relative to the mean motion. Symbols for different stellar classes are the same as in Figure 3. Each arrow marks the direction of the proper motion, with the length proportional to the speed. The blue arrows represent Gaia measurements, whereas the maroon arrows are for values from other data sets (see Section 2.2.1). The scale reference on the top right marks 10 mas yr$^{-1}$, which, in accord with the coordinate axes, corresponds to a motion of 6' in $3.5 \times 10^4$ yr.

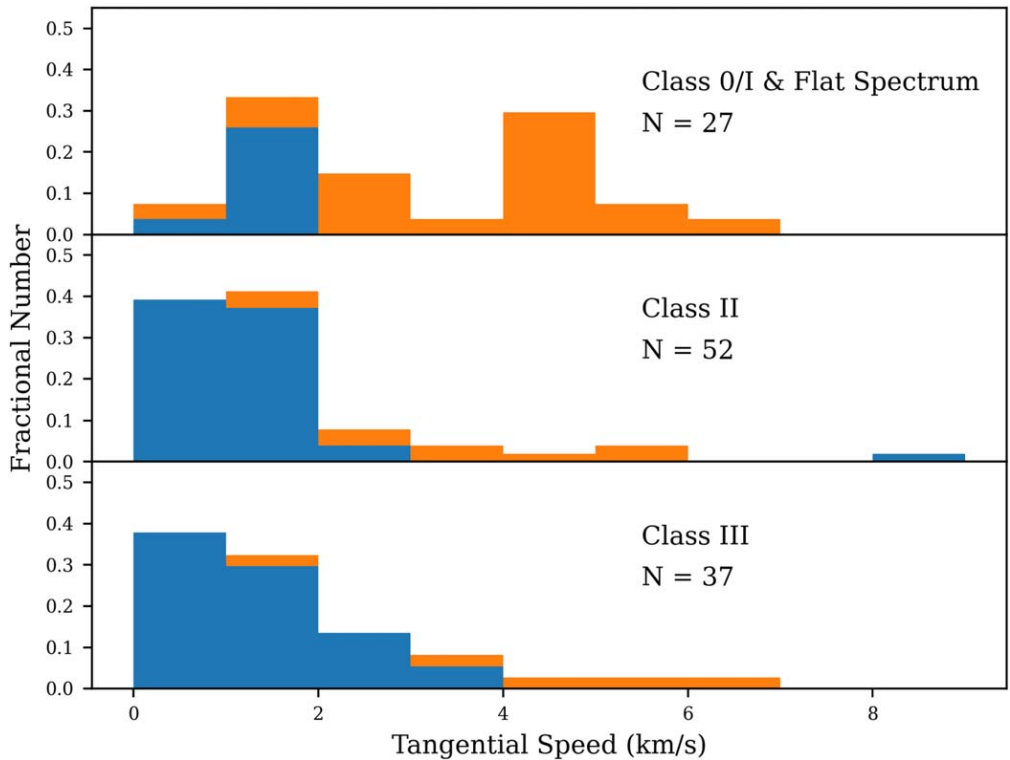

**Figure 7.** Distributions of relative speeds for YSOs of different classes. The blue bars represent the Gaia EDR3 measurements, whereas the orange bars denote the less-precise ground-based measurements (Monet et al. 2003; Ducourant et al. 2017).

White et al. (2015), using the $C^{18}O$ $J = 3-2$ emission, estimated a total molecular gas mass to be $\sim 515\,M_\odot$, consistent with those derived from other lines (e.g., $^{13}CO$ $J = 1-0$; Loren 1989; or $C^{18}O$ $J = 1-0$; Tachihara et al. 2000). White et al. (2015) measured a line width of 1.5 km s$^{-1}$, corresponding to an isotropic one-dimensional velocity dispersion of $\sim 0.64$ km s$^{-1}$. This value is comparable to the line-of-sight velocity dispersion between cores in Perseus (Kirk et al. 2007). Under the same assumption of virial equilibrium as in White et al. (2015), i.e., $\sigma_{vir}^2 = M\gamma G/5R$, where $\sigma_{vir}$ is the velocity dispersion, $M$ is the mass of molecular gas, $R$ is the radius of the cloud, $\gamma$ is the density profile factor (taken here to be 5/3 for a uniform sphere), and $G$ is the gravitational constant, the three-dimensional velocity dispersion with the adopted distance of 138 pc, assuming spherical symmetry, would be 0.96 km s$^{-1}$.

In comparison, our young stellar group has a three-dimensional velocity dispersion, each represented by fitting with a Gaussian, to be $2.42 \pm 0.33$ km s$^{-1}$ along the line of sight, $1.28 \pm 0.09$ km s$^{-1}$ in the R.A. direction, and $1.11 \pm 0.07$ km s$^{-1}$ in the decl. direction. The latter two, both in the sky plane, agree with each other, whereas the higher value and dispersion of the radial velocity can be attributed to larger observational uncertainties and possible stellar binarity. In any case, the values suggest a velocity dispersion of the young stars more than that of the natal gas, and also marginally greater than that expected for a system in virial equilibrium. For a sheet-like configuration stretching along our line of sight, the actual surface mass density, while uncertain, is largely overestimated by the measured column density, hence leading to a reduced velocity dispersion.

This increase of stellar velocity can be the consequence of dynamical evolution of the young cluster. Simulations of dynamics of young embedded clusters (e.g., Proszkow et al. 2009) demonstrate that as a system evolves from a subvirial state, the individual members tend to fall toward the center of the potential well and gain kinetic energy leading to a boosted velocity dispersion. Alternatively, the increased dispersion can also be due to the dispersal of gas by stellar feedback, leaving behind a shallower potential well and thus, a loosely bound supervirial cluster.

Moreover, as a star cluster ages, two-body relaxation among stellar members leads to mass segregation, i.e., lower-mass members would occupy progressively larger volume in space, vulnerable particularly for the lowest ones to subsequent ejection from the system, whereas the massive stars "sink" to the center. Because mass information is not readily available for our YSO objects, we investigated the kinematics on the spectral type of the YSOs listed in Allers & Liu (2020) and Esplin & Luhman (2020), as presented in Figure 8. YSOs earlier than the M type, while low in numbers, appear to have consistently tangential speeds up to 3 km s$^{-1}$ including the two Herbig Ae stars, SR 3 and IRAS 16245−2423. We note that the star IRAS 16245−2423 (YLS 46, or 2MASS J16273718−2430350) has uncertain spectral typing; Erickson et al. (2011), measuring a range of B5 to F2, adopted an A0 type. Those of M types or later seem to have noticeably a wider dispersion (up to $\sim 7$ km s$^{-1}$), notwithstanding the larger errors associated with non-Gaia proper-motion measurements (Monet et al. 2003; Ducourant et al. 2017) for the faint, late-type sources. All the Gaia data, however, indicate a consistent tangential velocity regardless of the spectral type. Given that only $\sim 17\%$ early-type stars have tangential speeds greater than $\sim 3$ km s$^{-1}$, the overall kinematics do not seem to vary greatly across different spectral types. Better astrometric data for the faint early-type sources is needed for further analysis.

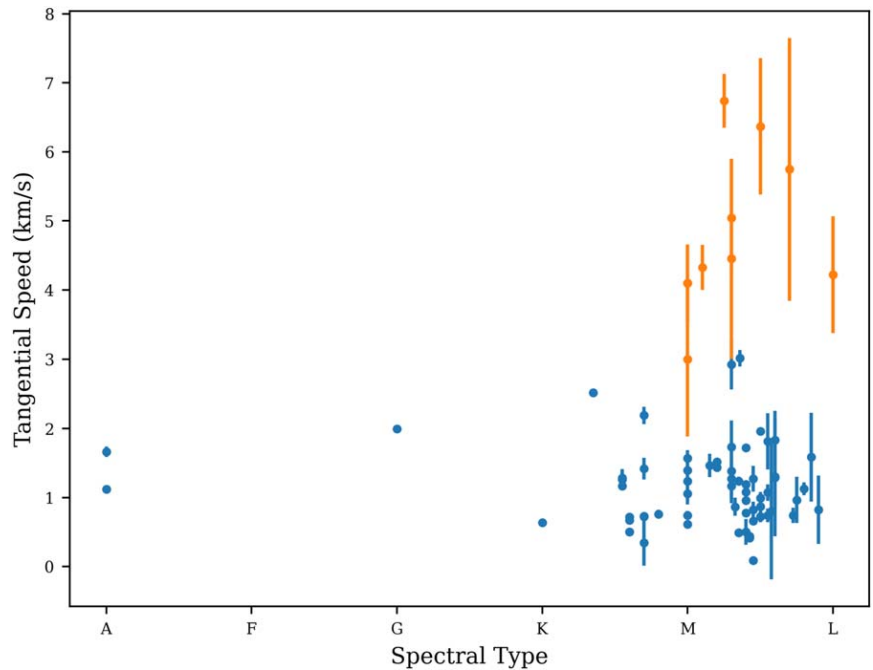

**Figure 8.** Tangential speed of YSOs vs. the spectral type. The blue symbol represents the Gaia EDR3 measurements, whereas the orange symbol marks the ground-based data (Monet et al. 2003; Ducourant et al. 2017), along with the error.

## 4. Conclusions

Using high-precision Gaia EDR3 proper-motion measurements of the young stars in L1688 of the Ophiuchus star-forming region, incorporating deep 850 $\mu$m JCMT/SCUBA2 image, in a field of ∼1300 square arcmin, we diagnose the interplay between starbirth and natal clouds to trace the star formation history in the complex. Our main results are:

1. With the JCMT/SCUBA-2 850 $\mu$m image, we derive the position and boundary of molecular clumps. The image, stacked from the JCMT Transient Survey aiming to detect possible brightness outbursts related to young stellar accretion, or magnetic activity, reaches ∼2.3 mJy per beam, deeper than previously reported in the literature.
2. The class I and flat-spectrum YSOs in the region are found to be closely associated with molecular clouds. The class II and class III are located preferentially further away from clouds, and are not randomly distributed in position; instead these young stars with aged disks show a structured spatial pattern that is devoid of molecular gas.
3. Stellar kinematics suggests Class II sources to have a comparable velocity dispersion as the Class III sources do. While data are of less precision for Class I and flat-spectrum sources, they do not seem to move slower in space. This suggests that the YSO populations are comoving in space and different evolutionary classes serve to diagnose the star formation history in the region.
4. The young stars exhibit segregated spatial distribution, with a loop of the most evolved young stars (class III), to those younger and associated with a part of cloud Oph-J, to the youngest near ongoing star-forming sites in the main cloud complex. This is consistent with a compression front from the OB stars located to the west and southwest.

We acknowledge the financial support of the grant MOST 109-2112-M-008-015-MY3 to carry out this study. The James Clerk Maxwell Telescope is operated by the East Asian Observatory on behalf of The National Astronomical Observatory of Japan, Academia Sinica Institute of Astronomy and Astrophysics of Taiwan, the Korea Astronomy and Space Science Institute, and Center for Astronomical Mega-Science (as well as the National Key R&D Program of China with No. 2017YFA0402700). Additional funding is provided by the Science and Technology Facilities Council of the United Kingdom and participating universities and organizations in the United Kingdom and Canada. The JCMT Transient Large Program has project codes M16AL001 and M20AL007. This research used the facilities of the Canadian Astronomy Data Centre operated by the National Research Council of Canada with the support of the Canadian Space Agency. This work has made use of data from the European Space Agency (ESA) mission Gaia (https://www.cosmos.esa.int/gaia), processed by the Gaia Data Processing and Analysis Consortium (DPAC; https://www.cosmos.esa.int/web/gaia/dpac/consortium).

*Facilities:* JCMT, Gaia, CTIO:2MASS, FLWO:2MASS, PS1, WISE, Spitzer, Herschel, MAST, ADS, CDS.

*Software:* astropy (Astropy Collaboration et al. 2013, 2018), scipy (Virtanen et al. 2020), pointpats (Rey et al. 2019), Starlink (Currie et al. 2014), MAKEMAP (Chapin et al. 2013).

### ORCID iDs

Aashish Gupta https://orcid.org/0000-0002-9959-1933
Wen-Ping Chen https://orcid.org/0000-0003-0262-272X

### References

Allers, K. N., & Liu, M. C. 2020, PASP, 132, 104401
Astropy Collaboration, Price-Whelan, A. M., Sipőcz, B. M., et al. 2018, AJ, 156, 123
Astropy Collaboration, Robitaille, T. P., Tollerud, E. J., et al. 2013, A&A, 558, A33

Bailer-Jones, C. A. L., Rybizki, J., Fouesneau, M., Demleitner, M., & Andrae, R. 2021, AJ, 161, 147
Berry, D. S. 2015, A&C, 10, 22
Cánovas, H., Cantero, C., Cieza, L., et al. 2019, A&A, 626, A80
Chambers, K. C., Magnier, E. A., Metcalfe, N., et al. 2016, arXiv:1612.05560
Chapin, E. L., Berry, D. S., Gibb, A. G., et al. 2013, MNRAS, 430, 2545
Chen, H., Myers, P. C., Ladd, E. F., & Wood, D. O. S. 1995, ApJ, 445, 377
Currie, M. J., Berry, D. S., Jenness, T., et al. 2014, in ASP Conf. Ser. 485, Astronomical Data Analysis Software and Systems XXIII, ed. N. Manset & P. Forshay (San Francisco, CA: ASP), 391
Ducourant, C., Teixeira, R., Krone-Martins, A., et al. 2017, A&A, 597, A90
Dunham, M. M., Allen, L. E., Evans, N. J. I., et al. 2015, ApJS, 220, 11
Dunham, M. M., Stutz, A. M., Allen, L. E., et al. 2014, in Protostars and Planets VI, ed. H. Beuther et al. (Tucson, AZ: Univ. of Arizona Press), 195
Erickson, K. L., Wilking, B. A., Meyer, M. R., Robinson, J. G., & Stephenson, L. N. 2011, AJ, 142, 140
Esplin, T. L., & Luhman, K. L. 2020, AJ, 159, 282
Frimann, S., Jørgensen, J. K., & Haugbølle, T. 2016, A&A, 587, A59
Gaia Collaboration, Brown, A. G. A., Vallenari, A., et al. 2021, A&A, 649, A1
Grasser, N., Ratzenböck, S., Alves, J., et al. 2021, A&A, 652, A2
Herczeg, G. J., Johnstone, D., Mairs, S., et al. 2017, ApJ, 849, 43
Holland, W. S., Bintley, D., Chapin, E. L., et al. 2013, MNRAS, 430, 2513
Johnstone, D., Herczeg, G. J., Mairs, S., et al. 2018, ApJ, 854, 31
Johnstone, D., Wilson, C. D., Moriarty-Schieven, G., et al. 2000, ApJ, 545, 327
Jørgensen, J. K., Johnstone, D., Kirk, H., et al. 2008, ApJ, 683, 822
Kirk, H., Johnstone, D., Di Francesco, J., et al. 2016, ApJ, 821, 98
Kirk, H., Johnstone, D., & Tafalla, M. 2007, ApJ, 668, 1042
Kumar, M. S. N., Arzoumanian, D., Men'shchikov, A., et al. 2021, A&A, 658, A114
Lada, C. J., & Lada, E. A. 2003, ARA&A, 41, 57
Ladjelate, B., André, P., Könyves, V., et al. 2020, A&A, 638, A74
Loren, R. B. 1989, ApJ, 338, 902
Mairs, S., Johnstone, D., Kirk, H., et al. 2016, MNRAS, 461, 4022
Mairs, S., Lane, J., Johnstone, D., et al. 2017, ApJ, 843, 55
Molinari, S., Schisano, E., Faustini, F., et al. 2011, A&A, 530, A133
Monet, D. G., Levine, S. E., Canzian, B., et al. 2003, AJ, 125, 984
Motte, F., Andre, P., & Neri, R. 1998, A&A, 336, 150
Nozawa, S., Mizuno, A., Teshima, Y., Ogawa, H., & Fukui, Y. 1991, ApJS, 77, 647
Ortiz-León, G. N., Loinard, L., Dzib, S. A., et al. 2018, ApJL, 869, L33
Pattle, K., Ward-Thompson, D., Kirk, J. M., et al. 2015, MNRAS, 450, 1094
Pilbratt, G. L., Riedinger, J. R., Passvogel, T., et al. 2010, A&A, 518, L1
Proszkow, E.-M., Adams, F. C., Hartmann, L. W., & Tobin, J. J. 2009, ApJ, 697, 1020
Retter, B., Hatchell, J., & Naylor, T. 2019, MNRAS, 487, 887
Rey, S., Kang, W., Shao, H., et al. 2019, pysal/pointpats: pointpats 2.1.0, Zenodo, doi:10.5281/zenodo.3265637
Ripley, B. D. 1976, J. Appl. Probab., 13, 255
Simpson, R. J., Nutter, D., & Ward-Thompson, D. 2008, MNRAS, 391, 205
Skrutskie, M. F., Cutri, R. M., Stiening, R., et al. 2006, AJ, 131, 1163
Stutz, A. M., & Gould, A. 2016, A&A, 590, A2
Sullivan, T., Wilking, B. A., Greene, T. P., et al. 2019, AJ, 158, 41
Tachihara, K., Mizuno, A., & Fukui, Y. 2000, ApJ, 528, 817
Torres, C. A. O., Quast, G. R., da Silva, L., et al. 2006, A&A, 460, 695
Virtanen, P., Gommers, R., Oliphant, T. E., et al. 2020, Nature Methods, 17, 261
Vrba, F. J. 1977, AJ, 82, 198
Vrba, F. J., Strom, S. E., & Strom, K. M. 1976, AJ, 81, 958
Werner, M. W., Roellig, T. L., Low, F. J., et al. 2004, ApJS, 154, 1
White, G. J., Drabek-Maunder, E., Rosolowsky, E., et al. 2015, MNRAS, 447, 1996
Wilking, B. A., Gagné, M., & Allen, L. E. 2008, in Star Formation in the ρ Ophiuchi Molecular Cloud, ed. B. Reipurth, Vol. 5 (San Francisco, CA: ASP), 351
Wilking, B. A., & Lada, C. J. 1983, ApJ, 274, 698
Williams, J. P., de Geus, E. J., & Blitz, L. 1994, ApJ, 428, 693
Wright, E. L., Eisenhardt, P. R. M., Mainzer, A. K., et al. 2010, AJ, 140, 1868